# Band parameters and hybridization in 2D semiconductor heterostructures from photoemission spectroscopy

Neil R Wilson[1,#,*], Paul V Nguyen[2,#], Kyle L Seyler[2], Pasqual Rivera[2], Alexander J Marsden[1], Zachary PL Laker[1], Gabriel C Constantinescu[3], Viktor Kandyba[4], Alexei Barinov[4], Nicholas DM Hine[1], Xiaodong Xu[2,5,*], David H Cobden[2,*]

[1]Department of Physics, University of Warwick, Coventry, CV4 7AL, UK
[2]Department of Physics, University of Washington, Seattle, Washington 98195, USA
[3]TCM Group, Cavendish Laboratory, University of Cambridge, 19 JJ Thomson Avenue, Cambridge CB3 0HE, UK
[4]Elettra - Sincrotrone Trieste, S.C.p.A., Basovizza (TS), 34149, Italy
[5]Department of Material Science and Engineering, University of Washington, Seattle, Washington 98195, USA
[#]These authors contribute equally to this work.
[*]Email: DHC (cobden@uw.edu), NRW (Neil.Wilson@warwick.ac.uk), XX (xuxd@uw.edu)

**Abstract**

**Combining monolayers of different two-dimensional (2D) semiconductors into heterostructures opens up a wealth of possibilities for novel electronic and optical functionalities. Exploiting them hinges on accurate measurements of the band parameters and orbital hybridization in separate and stacked monolayers, many of which are only available as small samples. The recently introduced technique of angle-resolved photoemission spectroscopy with submicron spatial resolution (μ-ARPES) offers the capability to measure small samples, but the energy resolution obtained for such exfoliated samples to date (~0.5 eV) has been inadequate. Here, we show that by suitable heterostructure sample design the full potential of μ-ARPES can be realized. We focus on $MoSe_2/WSe_2$ van der Waals heterostructures, which are 2D analogs of 3D semiconductor heterostructures. We find that in a $MoSe_2/WSe_2$ heterobilayer the bands in the K valleys are weakly hybridized, with the conduction and valence band edges originating in the $MoSe_2$ and $WSe_2$ respectively. There is stronger hybridization at the Γ point, but the valence band edge remains at the K points. This is consistent with the recent observation of interlayer excitons where the electron and hole are valley polarized but in opposite layers. We determine the valence band offset to be 300 meV, which combined with photoluminescence measurements implies that the binding energy of interlayer excitons is at least 200 meV, comparable with that of intralayer excitons.**

**Text**

An important direction in the growing field of 2D materials is the creation of artificial van der Waals heterostructures involving combinations of different monolayer materials, including semimetallic graphene, dielectric hexagonal boron nitride (hBN), and semiconductors such as transition metal dichalcogenides (TMDs).[1] Heterostructures of particular interest include graphene/hBN, whose Moire superlattice patterns produce unusual electronic structure;[2] graphene/TMD[3] and TMD/TMD, which show highly efficient photocurrent generation;[4–6] and graphene/hBN/TMD, which act as light emitting diodes.[7] Heterostructures of different 2D semiconductors could host exotic quantum phenomena such as superfluidity and Bose-Einstein condensation of excitons.[8] Ultrafast charge transfer between layers[9] and interlayer excitons ($X_I$) with the electron and hole in opposite layers,[10] have been observed in semiconducting TMD/TMD heterobilayers. They are also predicted to host rich valley physics[11] and indeed valley-polarized $X_I$ have been observed in aligned (small twist angle) samples.[12] A particularly promising combination is $WSe_2$ and $MoSe_2$, two isostructural semiconductors that are closely lattice-matched so they can be joined, stacked or alloyed with minimal strain and defects.[13]

Although progress has been rapid, the restrictions of optical and transport characterization leave many key questions open. For example, is a semiconductor heterobilayer still a direct-bandgap system with band edges at the K points? To what extent do the orbitals hybridize at the K and Γ points, and can one regard the bands at K simply as being those from isolated monolayers? What is the band offset? Is the interlayer exciton as strongly bound as excitons in the isolated monolayers are known to be? These questions illustrate the pressing need for direct and accurate measurements of the band parameters in order to establish authoritative basis of understanding of 2D heterostructures.

The technique of angle-resolved photoemission spectroscopy (ARPES) is well suited to determining 2D band structures. It has, for example, revealed bandgap opening[14] and effects of coexisting commensurate and incommensurate regions[15] in bilayer graphene, and yielded accurate bands in bulk layered semiconductors,[16–18] monolayer $MoSe_2$ epitaxially grown on graphite,[19] and graphene on bulk $MoS_2$.[20] However, until now high-resolution ARPES has been restricted by the photon beam profile to mm-scale samples, while μ-ARPES of exfoliated material has lacked the energy resolution needed to measure such quantities as band offsets and hybridization shifts.[21,22] Here we show that an order of magnitude better resolution can be achieved in μ-ARPES (<50 meV) by appropriate sample design, allowing accurate measurements of the relevant parameters in micron-scale 2D semiconductor heterostructures and also opening the way to studying the band structure in many other varieties of 2D device on a submicron spatial scale.

Fig. 1a is an optical image of a sample which illustrates our approach. In the center is an exfoliated flake of $WSe_2$ that has monolayer (1L), bilayer (2L) and multilayer (bulk) regions, with boundaries indicated by red dotted lines. Covering part of the flake is a graphene monolayer (G), outlined by a black dotted line, as indicated in the schematic cross-section in Fig. 1b. The flake was first picked up under the graphene on a polymer stamp (Supplementary Information S1), and the combination was then transferred onto a many-layer graphite flake exfoliated directly onto a p-doped silicon chip to serve as an atomically flat conducting substrate. Only very poor spectra could be obtained from layers placed directly on the silicon (Supplementary Information S2). The graphene cap allows the sample to be annealed at 400 °C in UHV to remove surface contamination without degrading the chalcogenides beneath it. Contamination between the van-der-Waals layers collects in blisters which consolidate on annealing to leave the majority of the interfaces between the monolayers atomically clean.[23]

The heterostructures were located in the μ–ARPES chamber by scanning photoemission microscopy (SPEM), in which a ~1 μm beam spot at 74 eV photon energy is scanned while collecting electrons with the spectrometer fixed (see Methods). At each pixel a spectrum was obtained, here integrated over a momentum range of ~ ±0.5 Å$^{-1}$ near Γ. Fig.1c is a SPEM map of the total emission in an energy window of ~ 3.5 eV below $E_F$. Fig. 1d shows sample spectra from the 1L, 2L and bulk regions. The 1L spectrum is alone in showing negligible photo-emission within 1 eV of $E_F$. In an SPEM map limited to this energy range (Fig. 1e) the 2L and bulk regions are visible but the monolayer is not. In addition, the peak in intensity shifts downwards monotonically in energy as the number of layers increases. As a result, a map of the peak energy vs position (Fig. 1f) shows contrast between 1L, 2L and bulk. Spectra from different points within a given region were identical (Supplementary Information S3), showing that the electronic properties were homogeneous, with no variation in work function as might be caused by contamination or compositional variation. Also, no drift in time due to charging was detected, showing that the graphite/silicon substrate makes a good electrical ground.

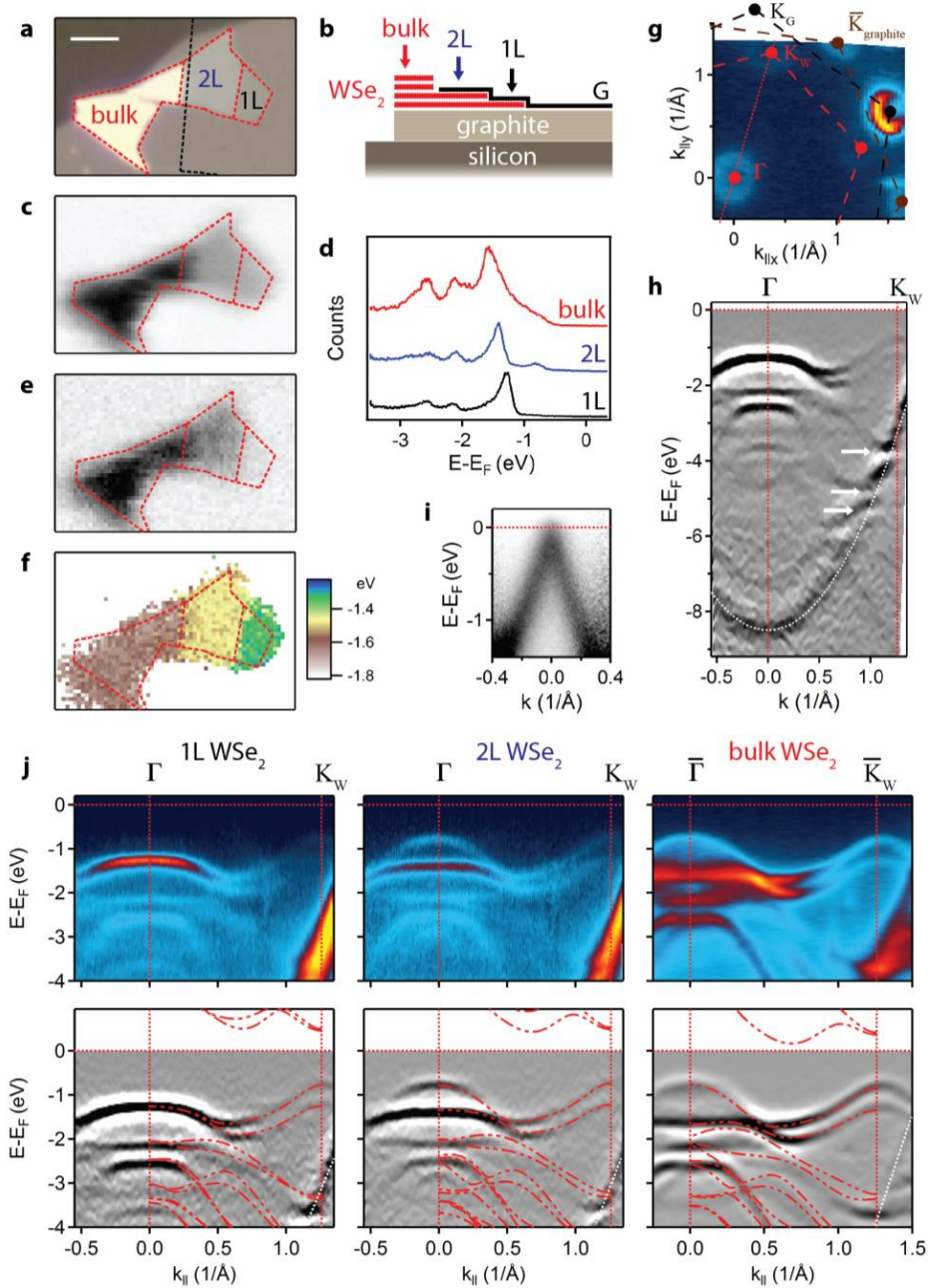

**Figure 1 | Micro-ARPES measurements on graphene-encapsulated WSe$_2$. a**, Optical image and **b**, schematic cross-section of an exfoliated WSe$_2$ flake with monolayer (1L), bilayer (2L) and bulk regions capped with monolayer graphene (G) and supported by a graphite flake on a doped silicon substrate. Scale bar is 5 μm. **c**, Scanning photoemission (SPEM) image of the same sample (integrated emission within ~ 3.5 eV of E$_F$ and ~ ±0.5 Å$^{-1}$ of Γ). **d,** Angle-integrated spectra from each region in **a**. **e**, SPEM image in the energy window -1 eV $< E - E_F < 0$. **f,** Map of the energy of maximum emission, showing contrast between 1L, 2L and bulk regions. **g,** Energy slice from full $E - k$ spectrum in the 1L region at $E - E_F = -0.9$ eV showing the Brillouin zone corners for the graphite (K$_{graphite}$), graphene (K$_G$) and WSe$_2$ (K$_W$). **h,** Momentum slice along Γ − K$_W$. Anticrossings between the graphene valence band (white dotted line) and the monolayer WSe$_2$ bands are indicated by white arrows. **i,** Momentum slice through the graphene K-point showing that $E_F$ is at the Dirac point. **j,** Comparison of $E - k$ spectra from 1L, 2L and bulk WSe$_2$ regions. The upper row shows total intensity, the lower is differentiated twice with respect to energy. The superposed red dashed lines are DFT calculations; the white dotted line indicates the graphene valence band.

Full $E$-**k** spectra were taken from points in the WSe$_2$ 1L, 2L and bulk regions. Fig. 1g shows an energy slice at $E - E_F = -0.9$ eV in the 1L region. Circles of valence band states can be seen around the Γ point and the corners (red, labelled K$_W$) of the hexagonal WSe$_2$ Brillouin zone. There are also circles around the corners of the graphene (black, K$_G$) and the graphite (brown, $\bar{\text{K}}_{\text{graphite}}$) zones, from which the relative orientations of the layers can be determined. The intensity from the graphite is an order of magnitude weaker than from the graphene, consistent with an inelastic mean free path of ~0.5 nm for ~70 eV electrons,[24] and underscoring the fact that the technique probes only the top few layers. Fig. 1h shows a momentum slice along Γ-K$_W$ in the 1L WSe$_2$ region. The graphene valence band (indicated by a white dotted curve) hybridizes with the WSe$_2$ bands producing avoided crossings (white arrows) similar to those previously reported for graphene on MoS$_2$.[20] Importantly, the crossings are >3 eV below $E_F$ and the WSe$_2$ bands nearer $E_F$ are not perturbed, as expected.[25] Fig. 1i shows a momentum slice through K$_G$ (on 1L WSe$_2$). The Dirac point $E_D$ coincides with the Fermi level (E$_F$ red dotted line) to within the measurement accuracy of $E_F$ of 50 meV, setting an upper bound on doping of the graphene at $4\pi(E_D - E_F)^2/(hv_F)^2$ ~10$^{11}$ carriers cm$^{-2}$ (with graphene Fermi velocity $v_F \approx 10^6$ ms$^{-1}$).

The 1L to 2L WSe$_2$ interface here constitutes a simple example of a 2D semiconductor heterostructure. Figure 1j compares Γ-K$_W$ slices in 1L, 2L and bulk regions. All features of the bands are well resolved, and density functional theory (DFT, overlaid red dashed lines) reproduces the upper valence band well with no adjustable parameters other than an energy offset chosen to match the uppermost measured band at Γ. The bilayer should be stacked as in the bulk 2H crystal. As expected,[26,27] in the monolayer the valence band maximum is at K (where the bandgap is known to be direct) and the bands near K are almost unchanged from monolayer to bulk. This is because the band edges at K have in-plane orbital character (W $5d_{xy}$ and $5d_{x^2-y^2}$) and hybridize very weakly between layers.[25,28] On the other hand, the bands at Γ have out-of-plane character (Se $4p_z$ and W $5d_{z^2}$) and we see strong hybridization effects. In the bilayer and bulk the valence band splits at Γ, with a higher-mass band about 0.25 eV below that in the monolayer and lower-mass band about 0.50 eV higher. In the bilayer the valence band edge is still at K, while in the bulk it moves to Γ.

We now turn to the heterostructures of primary interest, namely heterobilayers hosting interlayer excitons. Fig. 2a is an optical image of a sample with an MoSe$_2$ monolayer (green dashed line) partially overlapping a WSe$_2$ monolayer (red dashed line) to form a heterobilayer region (H, blue dotted line). The monolayers were intentionally aligned by identifying the crystal axes using polarization-resolved second harmonic generation[29–31] (Supplementary Information S4). Again there is a graphene cap and a graphite support. Fig. 2b shows angle-integrated spectra from a SPEM map acquired with the detector near Γ and E$_F$. The most intense peak in the MoSe$_2$ monolayer is ~200 meV lower than in the WSe$_2$, while in the H region there are two peaks that are shifted from those in the monolayers. As a result, a map of the energy of maximum intensity vs position (Fig. 2c) differentiates monolayer and H regions, while showing consistency within each region. Within 1 eV of E$_F$ there is no emission from the monolayers, but there is weak emission in the H region (Fig. 2d), as seen in the multilayer WSe$_2$ regions in Fig.1e. Full angle-resolved spectra were acquired at points in the monolayer and H regions. In constant-energy slices the locations of the K-points of the two monolayers coincide in momentum space (Supplementary Information S5), confirming a twist angle less than 1° (we do not know whether it was the 0° or 60° orientation) and consistent with their lattice constants differing by <1%.

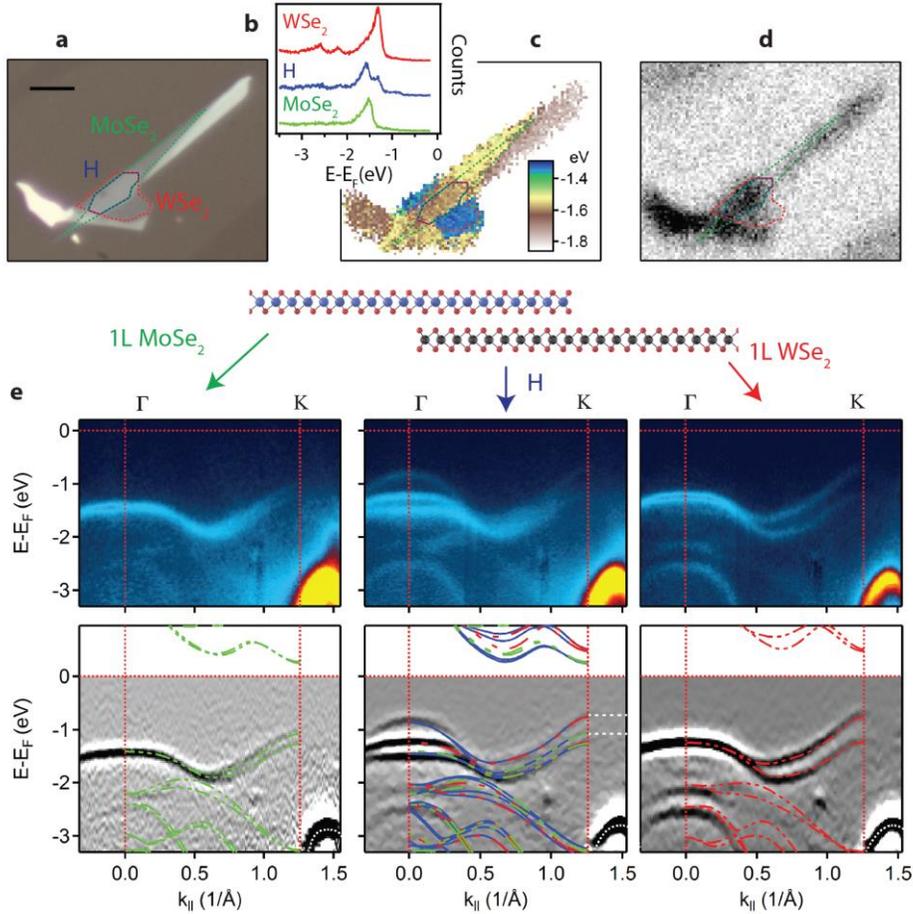

**Figure 2 | Photoemission measurements on a 2D heterostructure. a**, Optical image showing overlapping monolayer MoSe$_2$ on monolayer WSe$_2$ sheets in a heterobilayer region (H). Scale bar is 5 μm. Their boundaries are indicated with color-coded dashed lines. **b**, Angle-integrated spectra in each of the three regions. **c**, Map of the energy of maximum emission. **d**, SPEM image in the energy window -1 eV $< E - E_F <$ 0, showing contrast between H and the monolayers. **e**, Momentum slices along Γ-K in the three regions, with a cartoon of the structure above for reference. The superposed dashed lines on the twice differentiated data (bottom row) are DFT calculations for MoSe$_2$ monolayer (green), WSe$_2$ monolayer (red), and commensurate bilayer (blue). The graphene band is again indicated by a white dotted line. The white dashes indicate the valence band maxima in the MoSe$_2$ and WSe$_2$ monolayers and hence the valence band offset.

The variation in band structure between the three regions is revealed by the Γ−K momentum slices in Fig. 2e. The upper bands are again well matched by DFT (green and red dashed lines for MoSe$_2$ and WSe$_2$). The spin-orbit splitting is smaller in MoSe$_2$ than WSe$_2$ as expected, and we see that the valence band edge is lower in MoSe$_2$. In the H region (center panel), near K the bands are very similar to the bands in the monolayers, implying weak interlayer hybridization near K as in the WSe$_2$ bilayer. Near Γ there are two (more intense) bands again resembling those in the monolayers but shifted apart slightly, indicating moderate hybridization. Although the valence band edge remains at K, surprisingly there is an extra third (fainter) band at Γ with higher energy and smaller mass.

This presents a puzzle, since two bands cannot normally hybridize to give three (all should be spin-degenerate at Γ). We notice that the third band in the heterobilayer closely resembles the upper band in bilayer WSe$_2$ (Fig. 1j), which results from the stronger hybridization between commensurate layers. We also recall that when monolayers with different lattice constants are stacked, because of elastic energy any commensurate domains will have finite size as has been demonstrated for graphene on hBN.[32] For zero twist the scale of the domains is $a^2/\delta a$, where $a$ is the lattice constant and $\delta a$ is the difference. Here this scale is ~100 nm, less than the photon beam spot size. We can therefore understand the spectrum of the heterobilayer as being a superposition of spectra from a mixture of incommensurate domains in which hybridization is weak and commensurate domains in which hybridization is similar to that in 2H bilayer $WSe_2$. Indeed, DFT simulations of the commensurate heterobilayer reproduce the uppermost band at Γ (blue lines in the center panel), along with the slightly down-shifted lower band. Including the calculated dispersions of the isolated monolayers (also shown in green and red in the center panel) matches the three bands with reasonable agreement, excepting the small shifts at Γ mentioned earlier. These shifts, of order 100 meV, are reproduced by linear-scaling DFT[33] (Supplementary Information S6), and previous calculations have shown that their magnitude is roughly independent of twist angle for incommensurate structures.[34] This is because near Γ, unlike near K, the twist does not produce a crystal momentum mismatch that suppresses hybridization. As further evidence for the presence of commensurate and incommensurate domains, in another sample with an aligned monolayer of $WSe_2$ on a bilayer of $MoSe_2$ we observed four bands rather than three at Γ (Supplementary Information S7). Additionally, in a misaligned (by >5°) $WSe_2/WS_2$ heterobilayer we found only two bands at Γ (Supplementary Information S8), explained by the complete absence of commensuration in this case due to the combination of rotation and larger lattice mismatch (1.5 %).

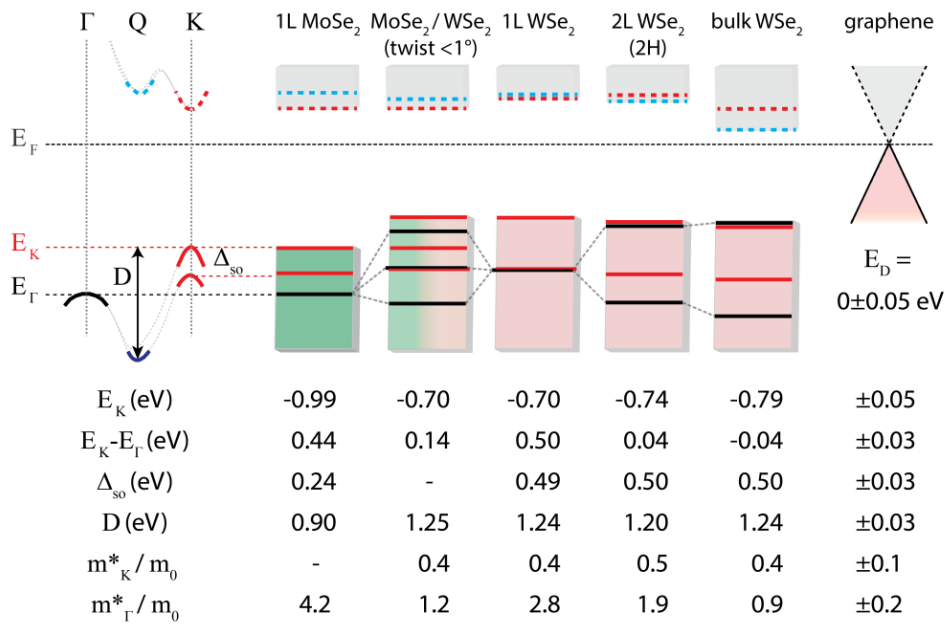

**Figure 3 | Summary of measured band parameters.** Left: schematic showing the definitions of parameters applicable for monolayers and aligned bilayers. Solid lines signify measured quantities; dotted lines DFT calculations. Main: graphical illustration of the shifts of homologous bands and hybridization effects. In both 2L $WSe_2$ and heterobilayer $MoSe_2/WSe_2$, hybridization is almost undetectable at K (red bands) but much larger at Γ (black bands). Table below: quantities determined by fitting the μ-ARPES spectra.

The values of key parameters extracted from the measurements are summarized in Fig. 3. They were found to be fully consistent across multiple samples and showed no dependence on the orientation of the graphene cap or graphite substrate. The spin-orbit splitting, $\Delta_{SO}$, at K is $0.49 \pm 0.03$ eV in WSe$_2$ and $0.24 \pm 0.03$ eV in MoSe$_2$, in agreement with the literature,[27] as are the effective masses of holes at Γ and K. In monolayer WSe$_2$ we find $E_K - E_\Gamma = 0.50 \pm 0.03$ eV, consistent with scanning tunneling spectroscopy results,[35] and in monolayer MoSe$_2$ we find $E_K - E_\Gamma = 0.44 \pm 0.03$ eV. We also record the valence band width, $D$, useful for comparison with band structure calculations.[27] As is well known, in both monolayer species the valence band edge is at K whereas in the bulk it is at Γ. In the heterobilayer we find that the valence band edge is also at K, higher than the maximum at Γ by $0.14 \pm 0.03$ eV. The valence band offset between WSe$_2$ and MoSe$_2$ monolayers is $\Delta_{VBO} = 0.30 \pm 0.03$ eV.

We cannot probe the conduction band, and the single-particle gaps have not been established incontrovertibly by other means, so we show the conduction band edges at K (red dashed line) and Q (blue dashed line) calculated by DFT. Although DFT underestimates these energies, the predicted variations within the family of materials and between different points in the Brillouin zone are more reliable.[27,28] The conduction band edge in the heterobilayer is predicted to remain at the K-point, which combined with our result implies that the band gap in H is direct.

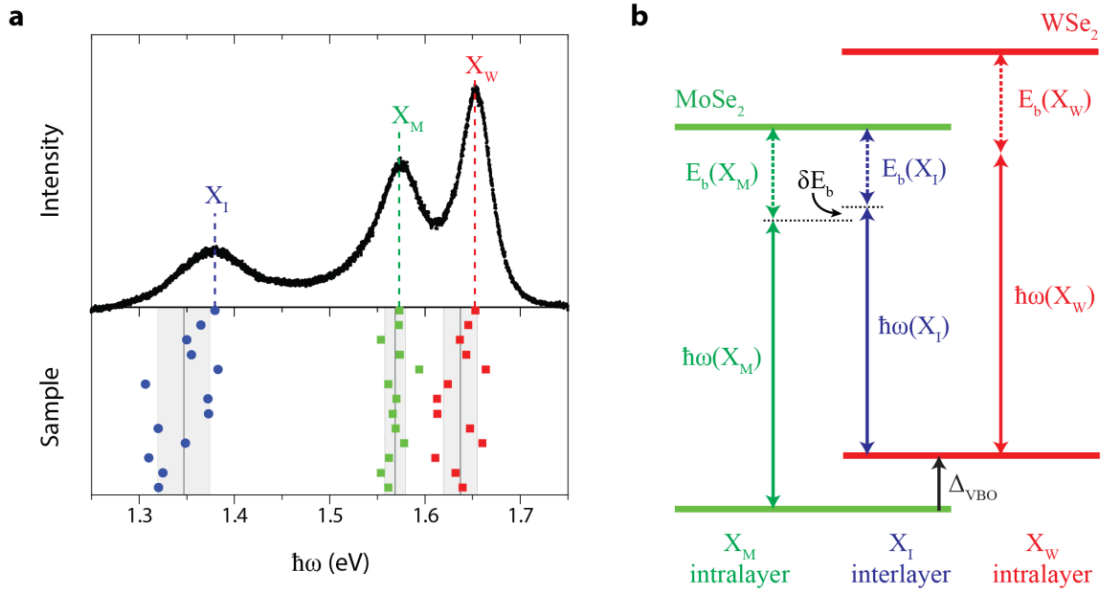

**Figure 4 | Photoluminescence and exciton binding in aligned MoSe$_2$/WSe$_2$ heterobilayers. a,** Above: representative PL spectrum showing peaks due to intralayer ($X_M$ and $X_W$) and interlayer ($X_I$) excitons (2.33 eV excitation at 20 $\mu$W). Below: peak positions for 13 samples, implying that the energy of $X_I$ is $220 \pm 20$ meV below that of $X_M$. **b,** Energy diagram showing the connection between the three exciton energies and the states derived from the MoSe$_2$ and WSe$_2$ conduction and valence bands at the K points.

We can now gain new insights by combining these ARPES results with the measurements of optical properties. Fig. 4a shows a photoluminescence spectrum from an aligned WSe$_2$/MoSe$_2$ heterobilayer sample at room temperature; below is a plot of the peak positions for 13 different samples. There are three peaks, whose origins are indicated schematically in Fig. 4b. $X_M$ and $X_W$ are the intralayer excitons formed from an electron and hole in bands from the same layer, either MoSe$_2$ or WSe$_2$, respectively. Their energies $\hbar\omega(X_M)$ and $\hbar\omega(X_W)$ are almost coincident with the corresponding valley excitons in the isolated monolayers, consistent with our finding that the band-edge states near the K-points hybridize little, and implying too that the binding energy of intralayer valley excitons in one layer is insensitive to the presence of the other layer. The third peak is the interlayer exciton, $X_I$. The small (~2%) variation of $\hbar\omega(X_I)$ between samples could be due to variations in substrate doping or twist angle.

According to Fig. 4b the energy difference between the intra- and interlayer excitons has two contributions, the difference in their binding energies, $\delta E_b = E_b(X_M) - E_b(X_I)$, and the valence band offset: $\hbar\omega(X_M) - \hbar\omega(X_I) = \Delta_{VBO} - \delta E_b$. Hence, by combining optical and ARPES measurements we can deduce that $\delta E_b = \hbar\omega(X_I) - \hbar\omega(X_M) + \Delta_{VBO}$. Averaging over the samples we get $\hbar\omega(X_M) - \hbar\omega(X_I) = 220 \pm 20$ meV, and using $\Delta_{VBO} = 300 \pm 30$ meV from above gives $\delta E_b = 80 \pm 40$ meV. The fact that the $X_I$ is more weakly bound than $X_M$ is not surprising because the electron and hole in different layers are on average further apart. The values of $E_b$ in similar monolayers in the literature range from ~300 to 700 meV,[36–43] with a report of 550 meV for MoSe$_2$.[36] We therefore infer that the interlayer binding energy $E_b(X_I) = E_b(X_M) - \delta E_b$ is at least ~200 meV. This is an order of magnitude larger than the binding energy of spatially indirect excitons in GaAs/AlGaAs double quantum wells.

It is clear that the technique of µ-ARPES provides invaluable information for understanding and realizing the potential of 2D semiconductor heterostructure devices. By combining it with sample design along the lines we have introduced here, it will be possible to determine accurately the local electronic structure and chemical potential in all manner of other 2D devices.

**Methods**

Samples were fabricated by exfoliation and dry transfer as detailed in Supplementary Information Sections 1 and 4.

µ-ARPES was performed at the Spectromicroscopy beamline of the Elettra light source, with linearly polarized radiation focused to a ~0.6 µm diameter spot by a Schwarzchild objective[44] and incident at 45° with respect to the sample. The energy and momentum resolution of the hemispherical electron analyzer were ~50 meV and ~0.03 Å$^{-1}$ respectively. SPEM maps were acquired over the energy range of the fixed detector (~3.5 eV), integrating over its angular range of ~15° (at 70 eV this is ~1.1 Å$^{-1}$). Samples were annealed at up to 700 K for 1 to 2 hours in UHV before measurement and the sample temperature during the measurements was 110 K.

The Quantum Espresso plane-wave DFT package[45] was used for calculations of individual materials and aligned heterostructures (Figs.1-3), including the spin-orbit interaction.[46] For simulations involving misaligned heterostructures, the ONETEP linear-scaling DFT code[33] was utilized. Further details are given in Supplementary Information S9.

**References**

1. Geim, a K. & Grigorieva, I. V. Van der Waals heterostructures. *Nature* **499,** 419–25 (2013).

2. Yankowitz, M. *et al.* Emergence of superlattice Dirac points in graphene on hexagonal boron nitride. *Nat. Phys.* **8,** 382–386 (2012).

3. Britnell, L. *et al.* Strong Light-Matter Interactions in Heterostructures of Atomically Thin Films. *Science (80-. ).* **340,** 1311–1314 (2013).

4. Cheng, R. *et al.* Electroluminescence and Photocurrent Generation from Atomically Sharp WSe 2 /MoS 2 Heterojunction p–n Diodes. *Nano Lett.* **14,** 5590–5597 (2014).

5. Furchi, M. M., Pospischil, A., Libisch, F., Burgdörfer, J. & Mueller, T. Photovoltaic Effect in an Electrically Tunable van der Waals Heterojunction. *Nano Lett.* **14,** 4785–4791 (2014).

6. Lee, C.-H. *et al.* Atomically thin p–n junctions with van der Waals heterointerfaces. *Nat. Nanotechnol.* **9,** 676–681 (2014).

7. Withers, F. *et al.* Light-emitting diodes by band-structure engineering in van der Waals heterostructures. *Nat. Mater.* **14,** 301–306 (2015).

8. Fogler, M. M., Butov, L. V. & Novoselov, K. S. High-temperature superfluidity with indirect excitons in van der Waals heterostructures. *Nat. Commun.* **5,** 1–5 (2014).

9. Hong, X. *et al.* Ultrafast charge transfer in atomically thin MoS2/WS2 heterostructures. *Nat. Nanotechnol.* **9,** 682–686 (2014).

10. Rivera, P. *et al.* Observation of long-lived interlayer excitons in monolayer MoSe2–WSe2 heterostructures. *Nat. Commun.* **6,** 6242 (2015).

11. Yu, H., Wang, Y., Tong, Q., Xu, X. & Yao, W. Anomalous Light Cones and Valley Optical Selection Rules of Interlayer Excitons in Twisted Heterobilayers. *Phys. Rev. Lett.* **115,** 187002 (2015).

12. Rivera, P. *et al.* Expansion of a Valley-Polarized Exciton Cloud in a 2D Heterostructure. (2016).

13. Huang, C. *et al.* Lateral heterojunctions within monolayer $MoSe_2$–$WSe_2$ semiconductors. *Nat. Mater.* **13,** 1–6 (2014).

14. Ohta, T. Controlling the Electronic Structure of Bilayer Graphene. *Science (80-. ).* **313,** 951–954 (2006).

15. Kim, K. S. *et al.* Coexisting massive and massless Dirac fermions in symmetry-broken bilayer graphene. *Nat. Mater.* **12,** 887–92 (2013).

16. Eknapakul, T. *et al.* Electronic structure of a quasi-freestanding $MoS_2$ monolayer. *Nano Lett.* **14,** 1312–6 (2014).

17. Riley, J. M. *et al.* Direct observation of spin-polarized bulk bands in an inversion-symmetric semiconductor. *Nat. Phys.* **10,** 835–839 (2014).

18. Latzke, D. W. *et al.* Electronic structure, spin-orbit coupling, and interlayer interaction in bulk MoS2 and WS2. *Phys. Rev. B* **91,** 235202 (2015).

19. Zhang, Y. *et al.* Direct observation of the transition from indirect to direct bandgap in atomically thin epitaxial MoSe2. *Nat. Nanotechnol.* **9,** 111–5 (2014).

20. Coy Diaz, H. *et al.* Direct Observation of Interlayer Hybridization and Dirac Relativistic Carriers in Graphene/$MoS_2$ van der Waals Heterostructures. *Nano Lett.* **15,** 1135–1140 (2015).

21. Jin, W. *et al.* Direct measurement of the thickness-dependent electronic band structure of MoS2 using angle-resolved photoemission spectroscopy. *Phys. Rev. Lett.* **111,** 1–5 (2013).

22. Yeh, P.-C. *et al.* Layer-dependent electronic structure of an atomically heavy two-dimensional dichalcogenide. *Phys. Rev. B* **91,** 1–6 (2015).

23. Kretinin, a. V. *et al.* Electronic Properties of Graphene Encapsulated with Different Two-Dimensional Atomic Crystals. *Nano Lett.* **14,** 3270–3276 (2014).

24. Seah, M. P. & Dench, W. a. Quantitative electron spectroscopy of surfaces: A standard data base for electron inelastic mean free paths in solids. *Surf. Interface Anal.* **1,** 2–11 (1979).

25. Komsa, H.-P. & Krasheninnikov, A. V. Electronic structures and optical properties of realistic transition metal dichalcogenide heterostructures from first principles. *Phys. Rev. B* **88,** 085318 (2013).

26. Liu, G.-B., Xiao, D., Yao, Y., Xu, X. & Yao, W. Electronic structures and theoretical modelling of two-dimensional group-VIB transition metal dichalcogenides. *Chem. Soc. Rev.* **44,** 2643–2663 (2015).

27. Kormányos, A. *et al.* k·p theory for two-dimensional transition metal dichalcogenide semiconductors. *2D Mater.* **2,** 022001 (2015).

28. Kang, J., Tongay, S., Zhou, J., Li, J. & Wu, J. Band offsets and heterostructures of two-dimensional semiconductors. *Appl. Phys. Lett.* **102,** 22–25 (2013).

29. Kumar, N. *et al.* Second harmonic microscopy of monolayer $MoS_2$. *Phys. Rev. B* **87,** 161403 (2013).

30. Li, Y. *et al.* Probing symmetry properties of few-layer MoS2 and h-BN by optical second-harmonic generation. *Nano Lett.* **13,** 3329–3333 (2013).

31. Malard, L. M., Alencar, T. V., Barboza, A. P. M., Mak, K. F. & De Paula, A. M. Observation of intense second harmonic generation from MoS2 atomic crystals. *Phys. Rev. B - Condens. Matter Mater. Phys.* **87,** 1–5 (2013).

32. Woods, C. R. *et al.* Commensurate–incommensurate transition in graphene on hexagonal boron nitride. *Nat. Phys.* **10,** 1–6 (2014).

33. Skylaris, C.-K., Haynes, P. D., Mostofi, A. A. & Payne, M. C. Introducing ONETEP: Linear-scaling density functional simulations on parallel computers. *J. Chem. Phys.* **122,** 084119 (2005).

34. Constantinescu, G. C. & Hine, N. D. M. Energy landscape and band-structure tuning in realistic MoS2 / MoSe2 heterostructures. *Phys. Rev. B* **91,** 195416 (2015).

35. Chiu, M. *et al.* Determination of band alignment in the single-layer MoS2/WSe2 heterojunction. *Nat. Commun.* **6,** 7666 (2015).

36. Ugeda, M. M. *et al.* Observation of giant bandgap renormalization and excitonic effects in a monolayer transition metal dichalcogenide semiconductor. **13,** 1091–1095 (2014).

37. He, K. *et al.* Tightly Bound Excitons in Monolayer WSe2. *Phys. Rev. Lett.* **113,** 026803 (2014).

38. Chernikov, A. *et al.* Exciton Binding Energy and Nonhydrogenic Rydberg Series in Monolayer WS2. *Phys. Rev. Lett.* **113,** 076802 (2014).

39. Wang, G. *et al.* Giant Enhancement of the Optical Second-Harmonic Emission of $WSe_2$ Monolayers by Laser Excitation at Exciton Resonances. *Phys. Rev. Lett.* **114,** 097403 (2015).

40. Klots, A. R. *et al.* Probing excitonic states in suspended two-dimensional semiconductors by photocurrent spectroscopy. *Sci. Rep.* **4,** 6608 (2014).

41. Ye, Z. *et al.* Probing excitonic dark states in single-layer tungsten disulphide. *Nature* **513,** 214–218 (2014).

42. Zhu, B., Chen, X. & Cui, X. Exciton Binding Energy of Monolayer WS2. *Sci. Rep.* **5,** 9218 (2015).

43. Zhang, C., Johnson, A., Hsu, C.-L., Li, L.-J. & Shih, C.-K. Direct Imaging of Band Profile in Single Layer MoS 2 on Graphite: Quasiparticle Energy Gap, Metallic Edge States, and Edge Band Bending. *Nano Lett.* **14,** 2443–2447 (2014).

44. Dudin, P. *et al.* Angle-resolved photoemission spectroscopy and imaging with a submicrometre probe at the SPECTROMICROSCOPY-3.2L beamline of Elettra. *J. Synchrotron Radiat.* **17,** 445–450 (2010).

45. Giannozzi, P. *et al.* QUANTUM ESPRESSO: a modular and open-source software project for quantum simulations of materials. *J. Phys. Condens. Matter* **21,** 395502 (2009).

46. Dal Corso, A. Pseudopotentials periodic table: From H to Pu. *Comput. Mater. Sci.* **95,** 337–350 (2014).

**Acknowledgments**

The Engineering and Physical Sciences Research Council (EPSRC), UK are acknowledged for support through studentships for AJM (EP/K503204/1) and ZPLL (EP/M506679/1). PVN was supported in part by a fellowship from the University of Washington Clean Energy Institute. Sample fabrication and data analysis were supported by DoE, BES, Materials Science and Engineering Division DE-SC0002197. Optical measurements and sample design were supported by NSF EFRI-1433496 and AFOSR FA9550-14-1-0277. NDMH and GC acknowledge the support of the Winton Programme for the Physics of Sustainability. Computing resources were provided by the Darwin Supercomputer of the University of Cambridge High Performance Computing Service. GC acknowledges the support of the Cambridge Trust European Scholarship.

# Supplementary information for 'Band parameters and hybridization in 2D semiconductor heterostructures from photoemission spectroscopy'

Neil R Wilson[1,#,*], Paul Nguyen[2,#], Kyle L. Seyler[2], Pasqual Rivera[2], Alexander J Marsden[1], Zachary PL Laker[1], Gabriel C. Constantinescu[3], Viktor Kandyba[4], Alexei Barinov[4], Nicholas DM Hine[1], Xiaodong Xu[2,5,*], David H Cobden[2,*]

[1]Department of Physics, University of Warwick, Coventry, CV4 7AL, UK
[2]Department of Physics, University of Washington, Seattle, Washington 98195, USA
[3]TCM Group, Cavendish Laboratory, University of Cambridge, 19 JJ Thomson Avenue, Cambridge CB3 0HE, UK
[4]Elettra - Sincrotrone Trieste, S.C.p.A., Basovizza (TS), 34149, Italy
[5]Department of Material Science and Engineering, University of Washington, Seattle, Washington 98195, USA
[#]These authors contribute equally to this work.
Email: DC (cobden@uw.edu), NW (Neil.Wilson@warwick.ac.uk), XX (xuxd@uw.edu)

## Contents

1. <u>Stamp fabrication process for encapsulated $WSe_2$ on silicon.</u>

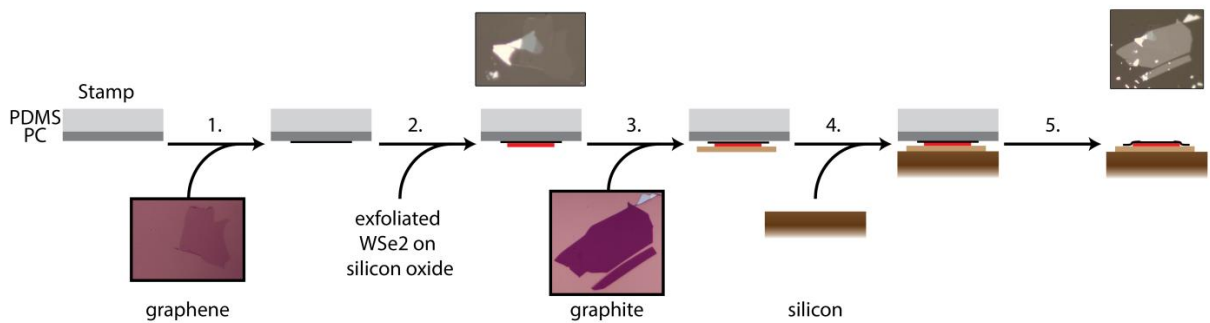

**Figure S1 | Schematic of the sample fabrication process for the $WSe_2$ sample in Fig. 1 of the main text.**

1. Graphene was exfoliated onto silicon oxide and a suitable flake identified by optical microscopy. The graphene flake was transferred onto a polycarbonate (PC) on polydimethylsiloxane (PDMS) stamp by dry transfer,[1] with a peak temperature of 105 °C to maximize graphene-to-PC adhesion. The stamp is on a glass slide mounted to a Maerzhaeuser Wetzlar (MW) SM 3.25 motorized micromanipulator controlled by joystick via an MW Tango controller.
2. $WSe_2$ was exfoliated onto silicon oxide and a suitable flake identified by optical microscopy. The graphene-on-stamp was aligned to the $WSe_2$ flake, heated to 90 °C against it, and then peeled off to remove the $WSe_2$ from the silicon oxide substrate, giving $WSe_2$-on-graphene-on-stamp.
3. A thin flake of graphite was exfoliated onto silicon oxide and identified by optical microscopy. The stamp was aligned to the graphite and dry transfer at 90 °C was again used to remove the graphite from the silicon oxide, giving a graphite-$WSe_2$-graphene stack on the stamp.
4. The stack and stamp were firmly placed on a doped silicon substrate.
5. The substrate was heated to 150 °C and the PDMS stamp peeled off, leaving the stack and PC on the silicon. The sample was then cleaned, removing the PC, by a solvent wash in chloroform, rinsed in IPA, and dried under $N_2$ gas followed by thermal annealing in an Argon (95%)/Hydrogen(5%) atmosphere at 400 °C for 2 hr.

2. Comparison of spectra on graphite and silicon

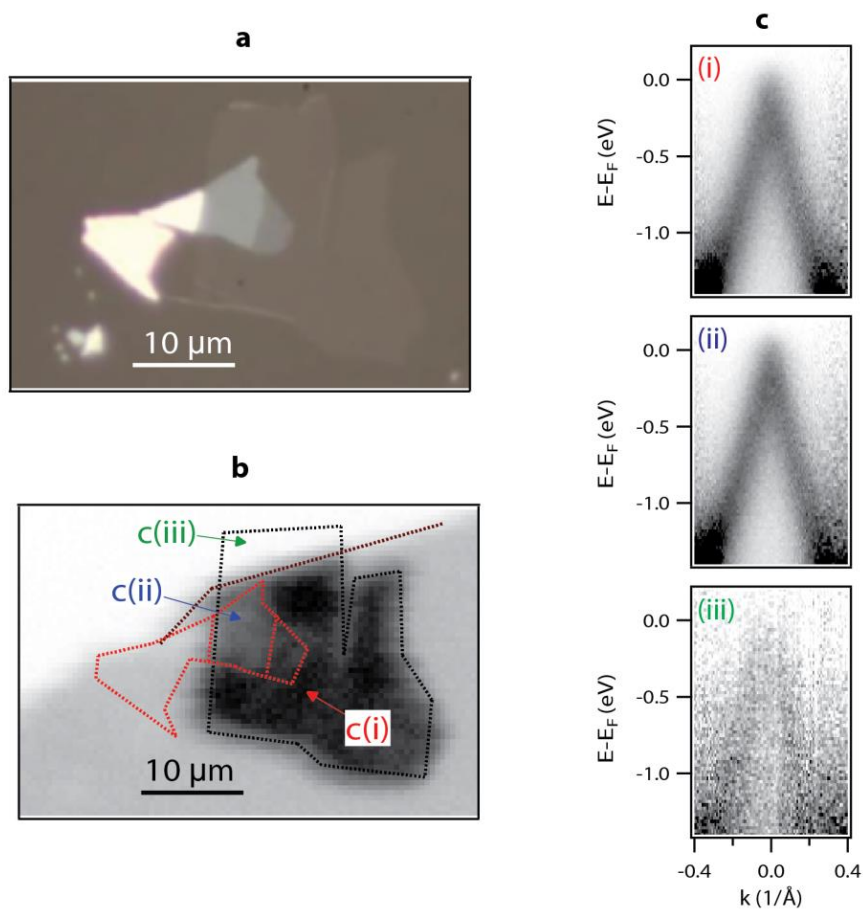

**Figure S2 | Importance of graphite substrate for clean spectra. a**, optical image of the graphene and WSe$_2$ on the stamp for the sample shown in Fig. 1 of the main text. **b**, SPEM map acquired at the K point for the graphene layer; the dashed lines outline the positions of the WSe$_2$ (red), graphite (brown) and graphene (black). **c**, energy-momentum slices across the K point of the graphene (i) on the graphene-on-graphite, (ii) on the graphene-on-bilayer WSe$_2$-on-graphite, and (iii) on the graphene-on-silicon (positions as marked in **b**). On silicon, the graphene bands are weak and diffuse. Similar results were found across all samples, with high resolution spectra only possible for monolayer / bilayer flakes on graphite not directly on silicon.

3. Uniformity of SPEM within 1L/2L/B regions on WSe$_2$

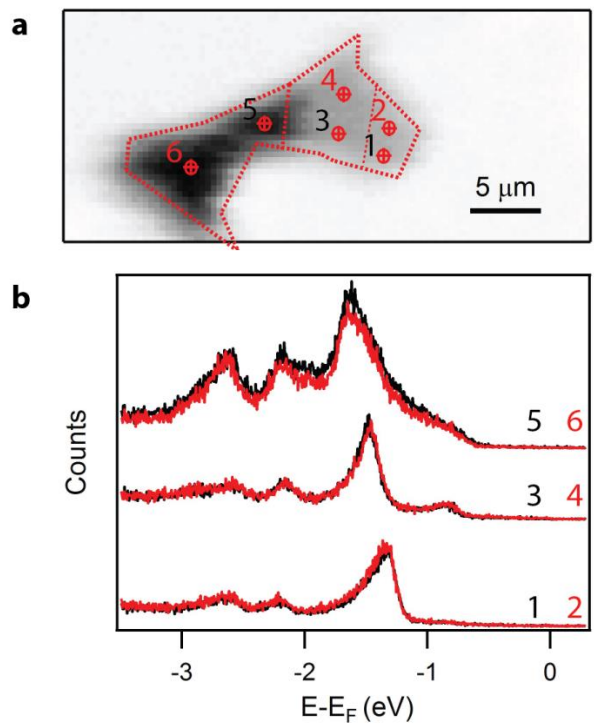

**Figure S3 | Demonstration of uniformity of photoemission within flakes. a**, SPEM image of the WSe$_2$ flake, as in Fig. 1c. **b** Spectra from within the monolayer, bilayer and multilayer regions of the WSe$_2$ from the positions marked in **a**, demonstrating the uniformity of behavior within each region.

4. Stamp fabrication process for encapsulated MoSe$_2$-on-WSe$_2$ heterobilayer

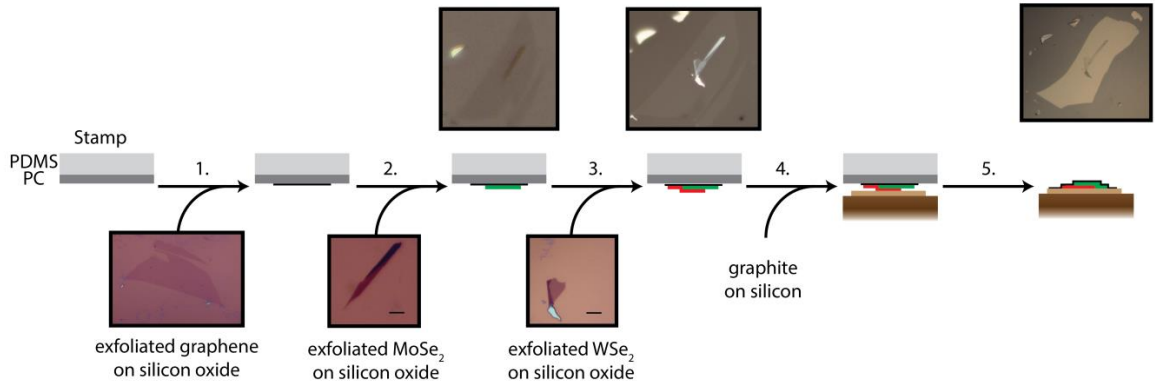

**Figure S4 | Schematic of the sample fabrication process for the MoSe$_2$-on-WSe$_2$ sample in Fig. 2 of the main text.** Scale bars are 5 μm.

1. Graphene was exfoliated onto silicon oxide and a suitable flake identified by optical microscopy. The graphene flake was transferred onto a polycarbonate (PC) on polydimethylsiloxane (PDMS) stamp by dry transfer,[1] with a peak temperature of 105 °C to maximize graphene-to-stamp adhesion. The stamp is on a glass slide which is taped to a rod mounted to a Maerzhaeuser Wetzlar (MW) SM 3.25 motorized micromanipulator controlled by joystick via an MW Tango controller.
2. MoSe$_2$ was exfoliated onto silicon oxide and a suitable flake identified by optical microscopy. The crystal axes were determined by room temperature linear-polarization-resolved second harmonic generation (SHG) at normal incidence with reflection geometry[2–4] with excitation at 1.4 μm. The graphene-on-stamp was aligned to the MoSe$_2$ flake, pressed against it, heated to 90 °C and then peeled off to remove the MoSe$_2$ from the silicon oxide substrate, giving MoSe$_2$-on-graphene-on-stamp.
3. WSe$_2$ was exfoliated onto silicon oxide and a suitable flake identified by optical microscopy. Its crystal axes were also determined by SHG with excitation at 1.5 μm. The MoSe$_2$-on-graphene-on-stamp was aligned to the WSe$_2$ flake, the flake was rotated to align the MoSe$_2$ and WSe$_2$ armchair axes within 2 degrees, and dry transfer at 90 °C was used to form the WSe$_2$-on-MoSe$_2$-on-graphene stack on the stamp.
4. A thin flake of graphite was exfoliated onto a doped silicon substrate and identified by optical microscopy and atomic force microscopy. The stamp was aligned, and the stack pressed firmly against the graphite.
5. The substrate was heated to 150 °C and the PDMS stamp peeled off, leaving the stack and PC on the silicon. The sample was then cleaned, removing the PC, by a solvent wash in chloroform, rinsed in IPA, and dried under N$_2$ gas followed by thermal annealing in an Argon(95%)/Hydrogen(5%) atmosphere at 400 °C for 2 hr.

5. ARPES constant energy slices showing relative orientations of MoSe$_2$ and WSe$_2$

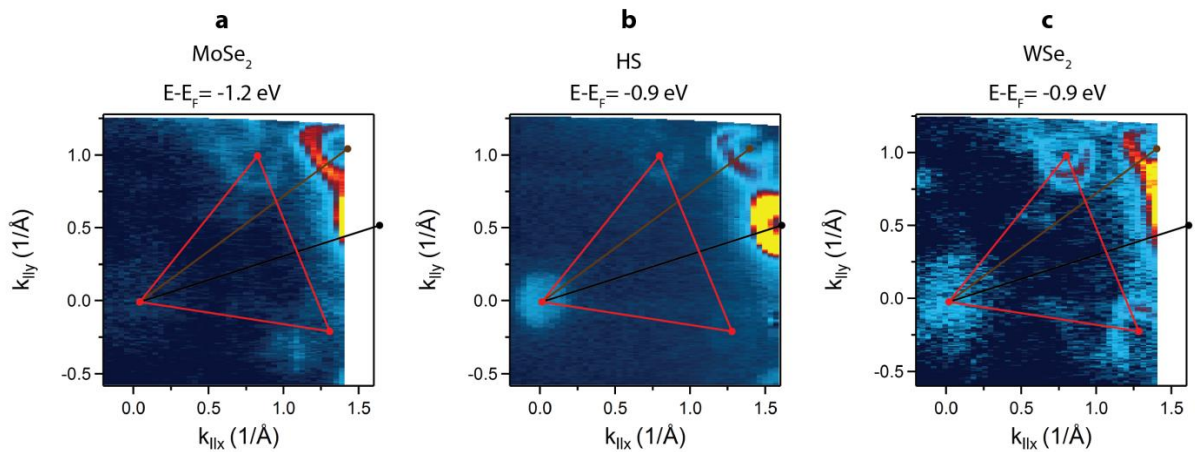

**Figure S5 | ARPES measurements of the orientations of the layers in the MoSe$_2$-on-WSe$_2$ heterostructure from Fig. 2 of the main text.** Constant energy slices with highlighted MX$_2$ Γ and K points (red circles connected by lines), graphene K point (black circle connected by line) and graphite K point (brown circle connected by line) from: **a** the MoSe$_2$ flake, **b** the heterostructure region and **c** the WSe$_2$ flake. The MX$_2$ Γ and K points were accurately found by fitting line profiles and coincided to within a rotation of 1° and a relative reciprocal spacing of 1%.

6. Linear scaling DFT calculations for misaligned MoSe$_2$-on-WSe$_2$ heterobilayers

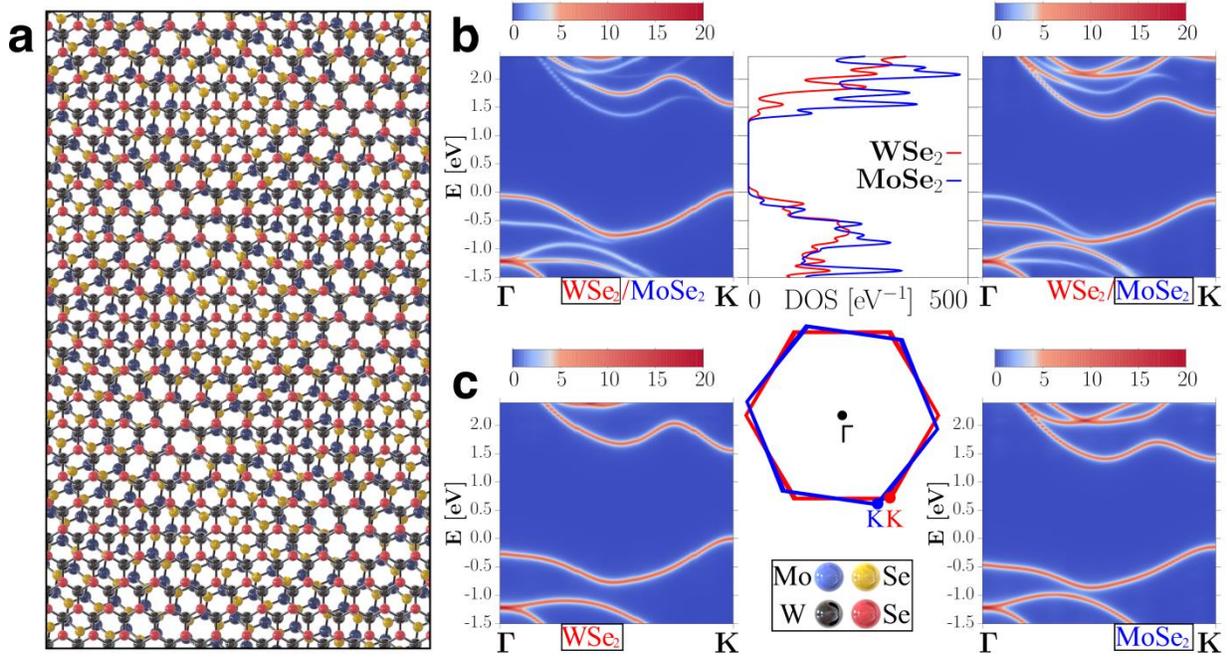

**Figure S6 | linear scaling DFT predictions of the bandstructure of the misaligned MoSe$_2$/WSe$_2$ interface**. Unfolded spectral function of the MoSe$_2$/WSe$_2$ heterostructure, projected on the WSe$_2$ monolayer (b-left) and MoSe$_2$ layer (b-right). The Γ-K directions are rotated by 8.21°. The spectral-function representation of the independent monolayers is also shown for WSe$_2$ (c-left) and MoSe$_2$ (c-right). The center inset of (b) shows the density of states projected onto the WSe$_2$ and MoSe$_2$ layers from the MoSe$_2$/WSe$_2$ interface. The energy reference is the energy of the valence band maximum (VBM) at K in the heterostructure.

Misaligned (rotated) MX$_2$ layers form coincidence cells too large for the capabilities of plane-wave DFT. Instead, linear-scaling DFT was used to gain insight into the effects of hybridization on band edge energies for incommensurate heterostructures. For full methodological details, see Section S9. We have previously shown that the energy landscape for misaligned MX$_2$ heterostructures is roughly independent of twist angle,[5] and similarly the hybridization induced shifts at Γ are consistent in magnitude. For this reason we considered the angle which resulted in the smallest supercell size; a simulation cell containing 873 atoms (432 for WSe$_2$, 441 for MoSe$_2$) for layers rotated by 8.21°, with strain < 1% (in the MoSe$_2$ layer). In order to observe the band structure effects of each layer in the presence of the other, we have calculated the unfolded spectral function, which was projected selectively on each of the component layers, as shown in Fig. S6b and as described in detail in our previous works.[5]

Comparison between the band structures of the independent monolayers and the unfolded spectral-functions of the corresponding layers in the heterostructure shows low-spectral weight band intrusions from one monolayer into the other upon stacking. Moreover, the valence band maximum (VBM) at Γ of WSe$_2$ is raised by 202 meV, while the MoSe$_2$ VBM is lowered by 67 meV giving an increase in separation of ~ 250 meV. The experimental results show qualitatively similar behavour, with the bands shifting in the same direction and with a similar magnitude of increase in separation (experimentally ~ 100 meV) to the DFT predictions. Note that currently the linear scaling DFT approach adopted here does not include spin-orbit interactions – these are not expected to significantly alter the band structure at Γ but do change the band structure at K.

7. ARPES of encapsulated MoSe$_2$-on-WSe$_2$ with heterotrilayer regions

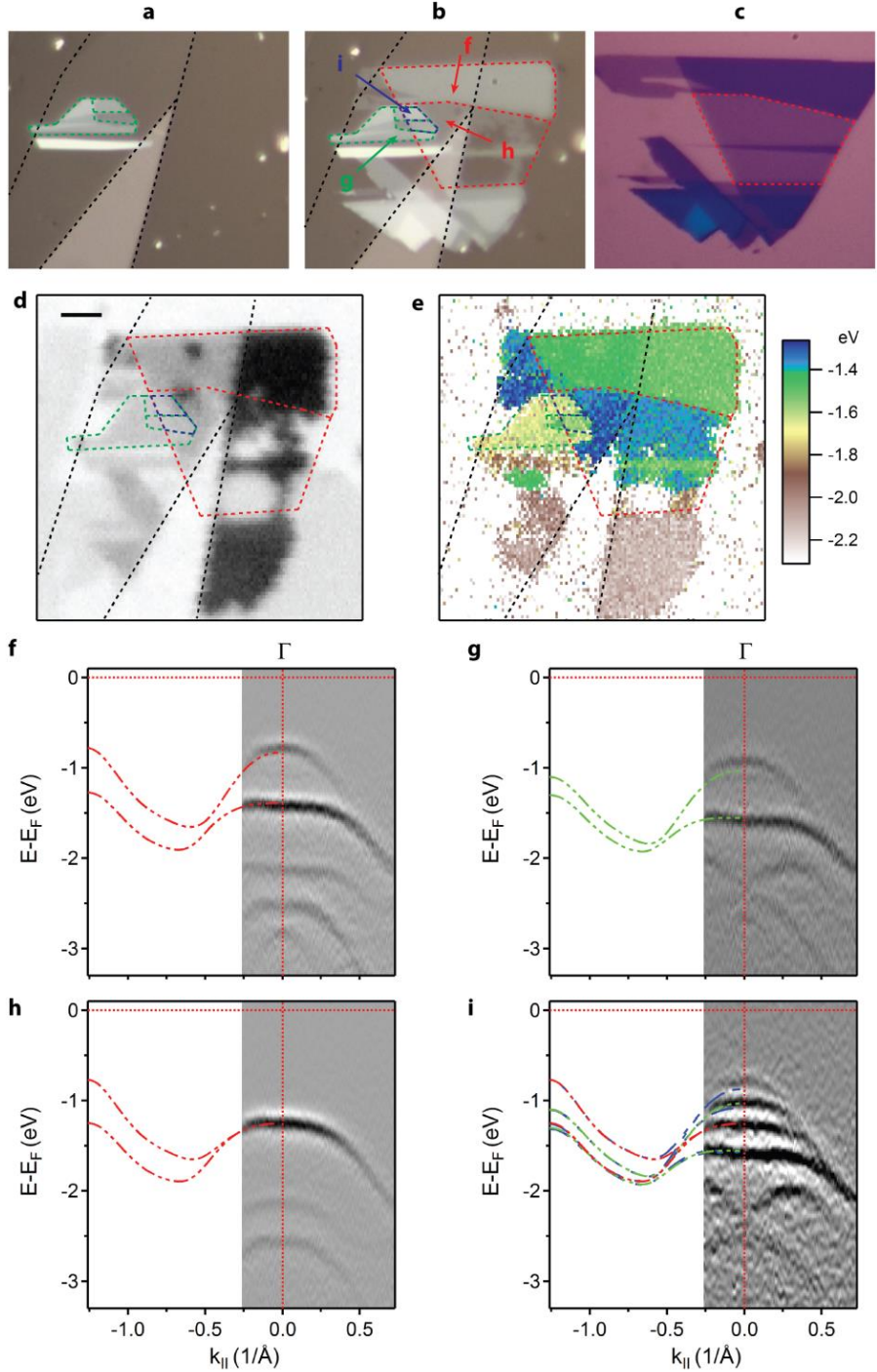

**Figure S7 | µ-ARPES of encapsulated MoSe$_2$-on-WSe$_2$ with heterotrilayer regions.** Optical images of: **a** the MoSe$_2$ flake (outlined by dashed green line) on graphene (outlined by dashed black line) on the stamp prior to transfer with the heterostructure region outlined in blue; **b** the WSe$_2$ (outlined by dashed red line), MoSe$_2$ and graphene on the stamp, and **c** the exfoliated WSe$_2$ flake on silicon oxide prior to transfer. **d** Integrated SPEM map at Γ near E$_F$; scale bar is 5 µm. **e** Corresponding map of the energy of maximum emission. The dispersion around Γ is shown in **f-i** from points in the bilayer WSe$_2$, bilayer MoSe$_2$, monolayer WSe$_2$, and the heterotrilayer (bilayer MoSe$_2$ on monolayer WSe$_2$) regions respectively with corresponding DFT calculations

overlaid: for the heterotrilayer the independent layers (monolayer WSe$_2$ red dashed, bilayer MoSe$_2$ green dashed) and commensurate heterotrilayer (blue dashed) are both shown. Unfortunately for this sample the drift in position during acquisition was too quick to acquire full $E - \bm{k}$ spectra from each region, so these dispersions are not from high symmetry directions.

## 8. Band structure of monolayer WSe$_2$-on-WS$_2$

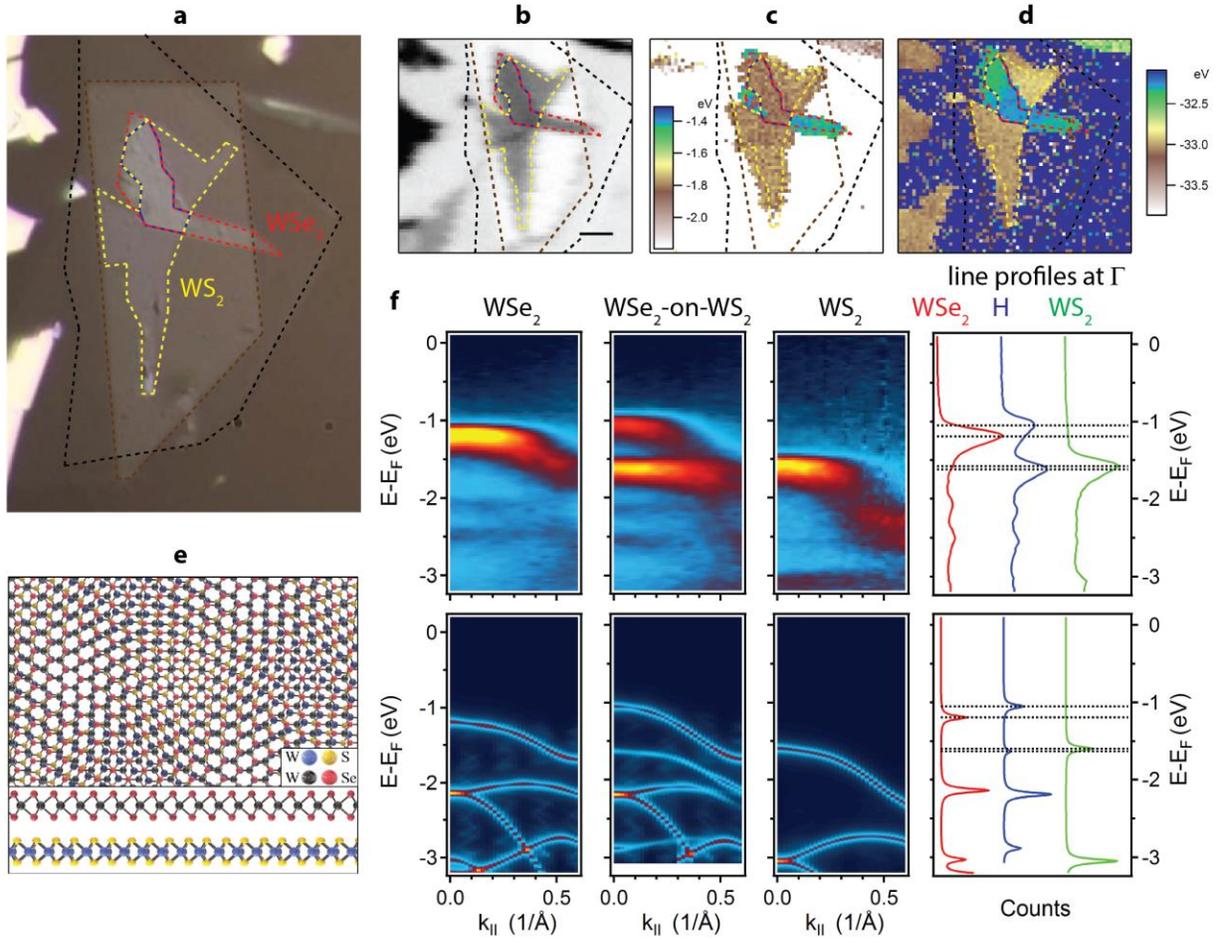

**Figure S8 | band structure of monolayer WSe$_2$-on-WS$_2$. a** Optical image of the graphene (outlined by black dashed line) on monolayer WSe$_2$ flake (red dashed line) on monolayer WS$_2$ (yellow dashed line) on graphite (black dashed line) with the heterostructure region outlined in blue. **b** Integrated SPEM map at Γ near E$_F$; scale bar is 5 µm. **c** Corresponding map of the energy of maximum emission. In the heterostructure region the WS$_2$ valence band at Γ has lower mass, giving a more pronounced peak in angle integrated spectra and hence is the more intense peak despite the WS$_2$ being underneath. **d** SPEM map of the peak energy vs position in the W 4f core level region (at Γ), this shows that the WSe$_2$ layer is on top. **e** Atomic cartoon of the WSe$_2$/WS$_2$ twisted bilayer. **f** Dispersion around Γ from the regions as marked: the top row are the experimental results and the lower row LS-DFT predictions, on the right are line profiles through the bands at Γ for the WSe$_2$ (red), WS$_2$ (green) and the twisted bilayer (blue) showing that the hybridization induced shifts in band edges predicted by the LS-DFT are well matched to the µ-ARPES data.

A twisted bilayer of WSe$_2$ on WS$_2$, encapsulated between graphene and graphite, was also investigated. As shown in Fig. S8, in the heterostructure region only two bands were seen at Γ, shifted slightly from the bands in the independent monolayers (WSe$_2$ shifted up by 150 meV and WS$_2$ shifted down by 60 meV). The spectra from this sample were too diffuse to clearly resolve the dispersions at K and so these dispersions are not definitively along a high symmetry direction.

LS-DFT was again used to gain insight into the band shifts. The supercell contained 762 atoms (363 for $WSe_2$, 399 for $WS_2$), with a 4.31° rotation between layers, giving a strain of < 1% (in the $WS_2$ layer). Unfolded spectral functions for the independent layers, and for the heterostructure region, are given in Fig.S8: it is clear that when $WS_2$ and $WSe_2$ are stacked, there is evidence of hybridisation, as bands intrude from one monolayer into the other. In the heterostructure region the mass of the $WS_2$ band at $\Gamma$ increases and its energy shifts down by 20 meV, whilst the $WSe_2$ band increases in energy by 200 meV and its mass decreases: both observations are consistent with the experimental results. The LS-DFT thus shows that the observed changes in the band structure in the heterostructure are fully consistent with the expected effects of hybridization between layers.

From the experimental results we can also measure the valence band offset at $\Gamma$ to be 0.38 ± 0.03 eV for the $WSe_2/WS_2$ heterostructure ($WSe_2$ higher in energy). Although we could not resolve the bands at K for this sample, we can combine the experimental band positions at $\Gamma$ with plane wave DFT valence band structures for $WS_2/WSe_2$ (including spin-orbit interactions) to predict the valence band offset at K to be 0.6 eV ($WSe_2$ higher in energy).

9. DFT methodology

Plane-wave DFT: for calculations involving individual materials and aligned heterostructures, the Quantum Espresso[6] plane-wave DFT package was used. The ultrasoft atomic datasets of Garrity et al[7] were used for structural calculations, and the optB88-vdW functional[8] was employed, due to its previous success in describing interlayer interactions in 2D materials.[5] The structures were optimized until forces were smaller than $10^{-4}$ Ry / Bohr for monolayers, and $5\times10^{-4}$ Ry / Bohr for bilayers and bulk, while stresses were required to be smaller than 0.05 GPa. Subsequently, the bandstructures were calculated using the high-accuracy fully-relativistic PAW potentials of Dal Corso,[9] such that spin-orbit interaction was included. We used a 12x12 in-plane k-point sampling grid (with 4 out-of-plane k-points for the bulk), an 800 eV plane-wave energy cutoff, and an 8000 eV charge density cutoff. The simulation cell height was 30.0 Å, to avoid interaction between periodic images. All these parameters were determined to be sufficient for very good convergence of structural and electronic properties.

Linear-scaling DFT: we utilized the ONETEP code,[10] which uses an efficiently-parallelised linear-scaling formalism[11] based around representation of the single-electron density matrix via in-situ optimized local orbitals and sparse matrices. Once again we used the optB88-vdW functional and a kinetic-energy cutoff of 800 eV. ONETEP does not currently have the ability to include spin-orbit coupling, the projector-augmented wave (PAW) method was employed, with atomic datasets exactly equivalent to the ultra-soft pseudopotential (USPP) datasets used for the geometry optimisations in the plane-wave DFT calculations described above. The Mo and W atoms both contained 14 valence electrons ($4s^2, 4p^6, 4d^5, 5s^1$ for Mo, $5s^2, 5p^6, 5d^4, 6s^2$ for W), while S and Se contained only 6 valence electrons ($3s^2, 3p^4$ for S, $4s^2, 4p^4$ for Se).

ONETEP uses a nested-loop optimization scheme in which an outer loop optimizes the form of the local orbitals, while an inner loop optimizes the density matrix for fixed local orbitals. The flexibility provided by in-situ optimization means that it is possible to use relatively small number of local orbitals and retain systematically controllable accuracy equivalent to the plane-wave approach. In this case we used 13 non-orthogonal Wannier functions (NGWFs) for W and Mo (10 for the valence electrons, 3 allowing for additional polarisation) and 9 for S and Se (4 for the valence electrons, 5 for additional variational freedom). All NGWFs were chosen to have a large cut-off radius (13.0 bohr), and the convergence criterion was that the root mean square of the NGWF gradient be smaller than $2\times10^{-6}$. For each NGWF optimisation step, 8 self-consistent density-kernel iterations were performed. Truncation of the density kernel was not necessary for the system sizes employed. Geometry optimisation was performed by relaxing the internal atomic coordinates until the forces[12] were below 0.1 eV / Å. The supercell was constructed by first determining the coincidence cells of the over-lapping rotated monolayers, allowing a maximum of 1% strain. Spectral functions were calculated by unfolding supercell eigenstates into the primitive cells of each layer, as described in previous work.[5]

10. References

1. Zomer, P. J., Guimarães, M. H. D., Brant, J. C., Tombros, N. & van Wees, B. J. Fast pick up technique for high quality heterostructures of bilayer graphene and hexagonal boron nitride. *Appl. Phys. Lett.* **105,** 013101 (2014).

2. Kumar, N. *et al.* Second harmonic microscopy of monolayer MoS$_2$. *Phys. Rev. B* **87,** 161403 (2013).

3. Li, Y. *et al.* Probing symmetry properties of few-layer $MoS_2$ and h-BN by optical second-harmonic generation. *Nano Lett.* **13,** 3329–3333 (2013).

4. Malard, L. M., Alencar, T. V., Barboza, A. P. M., Mak, K. F. & De Paula, A. M. Observation of intense second harmonic generation from $MoS_2$ atomic crystals. *Phys. Rev. B - Condens. Matter Mater. Phys.* **87,** 1–5 (2013).

5. Constantinescu, G. C. & Hine, N. D. M. Energy landscape and band-structure tuning in realistic $MoS_2$ / $MoSe_2$ heterostructures. *Phys. Rev. B* **91,** 195416 (2015).

6. Giannozzi, P. *et al.* QUANTUM ESPRESSO: a modular and open-source software project for quantum simulations of materials. *J. Phys. Condens. Matter* **21,** 395502 (2009).

7. Garrity, K. F., Bennett, J. W., Rabe, K. M. & Vanderbilt, D. Pseudopotentials for high-throughput DFT calculations. *Comput. Mater. Sci.* **81,** 446–452 (2014).

8. Klimeš, J., Bowler, D. R. & Michaelides, A. Chemical accuracy for the van der Waals density functional. *J. Phys. Condens. Matter* **22,** 022201 (2010).

9. Dal Corso, A. Pseudopotentials periodic table: From H to Pu. *Comput. Mater. Sci.* **95,** 337–350 (2014).

10. Skylaris, C.-K., Haynes, P. D., Mostofi, A. A. & Payne, M. C. Introducing ONETEP: Linear-scaling density functional simulations on parallel computers. *J. Chem. Phys.* **122,** 084119 (2005).

11. Hine, N. D. M. *et al.* Accurate ionic forces and geometry optimization in linear-scaling density-functional theory with local orbitals. *Phys. Rev. B* **83,** 195102 (2011).

12. Wilkinson, K. A., Hine, N. D. M. & Skylaris, C.-K. Hybrid MPI-OpenMP Parallelism in the ONETEP Linear-Scaling Electronic Structure Code: Application to the Delamination of Cellulose Nanofibrils. *J. Chem. Theory Comput.* **10,** 4782–4794 (2014).